\documentclass[numreferences]{kluwer}
\usepackage{epsfig,graphicx,klucite}

\def\beq{\begin{eqnarray}}
\def\eeq{\end{eqnarray}}

\begin{document}
\begin{article}

\begin{opening}
\title{Chaos and Brane-worlds\footnote{Based on a work done in collaboration with
J.D. Barrow.}}
\author{Sigbj\o rn Hervik\email{\uppercase{S}.\uppercase{H}ervik@damtp.cam.ac.uk}} \institute{
DAMTP, Centre for Mathematical Sciences, Cambridge University, \\ Wilberforce Rd., Cambridge CB3 0WA, UK}

\begin{motto}
``My brain is always chaotic.''
\end{motto}

\begin{abstract}
The early time behaviour of brane-world models is analysed in the presence of
anisotropic stresses. It is shown that that the initial singularity
cannot be isotropic, unless there is also  an isotropic fluid stiffer
than radiation present. Also, a magnetic Bianchi type I
brane-world is analysed in detail. It is known that the Einstein equations for the magnetic Bianchi type
I models are in general oscillatory and are believed to be chaotic, but in
the brane-world model this chaotic behaviour does not seem to be
possible.
\end{abstract}


\end{opening}
\section{Introduction}
The classical Einstein equations  have been shown to have  a very peculiar
feature, namely chaos
\cite{Spokoiny:1981fs,Belinsky:1970ew,jb2,jb3,cb,dem,DH2000,dh2}. As
the initial singularity is approached some cosmological
solutions seem to oscillate chaotically. This type of  behaviour
in the general relativistic vacuum Bianchi type VIII and IX models
have been well studied \cite{hbc}, and chaotic behaviour for the general
relativistic magnetic Bianchi type I model has been conjectured
\cite{LeBlanc}. 

The question we will address here is the following: Is the same
chaotic behaviour present in the relatively newly proposed brane-world
models? \cite{Randall:1999vf,Randall:1999ee}. Earlier investigations
have shown that the brane-world
models are more sensitive to the matter content than its general
relativistic (GR) counterpart,
especially at high energies \cite{BDL}. The reason for this is that the Friedmann
equation on the brane contains quadratic terms in the energy-density
compared to the usual linear term in ordinary GR cosmology.  

An important set of solutions in the study of the classical general relativistic cosmologies
 are the ones found by Kasner \cite{Kasner}. These
 are vacuum Bianchi type I solutions with  metric
\beq
ds^2=-dt^2+t^{2p_1}dx^2+t^{2p_2}dy^2+{t^{2p_3}dz^2 }
\eeq
where $\sum_ip_i=\sum_ip^2_i=1$.  They correspond to highly
 anisotropic universes and
have a very special role in GR cosmology. For most types of matter
(except, for example, a stiff fluid), all spatially homogeneous universe
 models will -- as we approach the initial singularity
\begin{itemize}
\item asymptote to one of the Kasner solutions, \textit{or}
\item have \textbf{chaotic} oscillations between different Kasner
epochs.
\end{itemize} (c.f. \cite{DynSys,BS})
In particular, in the latter case the Kasner
solutions are unstable in the past while in the former they are
stable.  Hence, in GR cosmology the initial
singularity will in general be anisotropic. 

\begin{figure}
\centerline{\includegraphics[height=4cm]{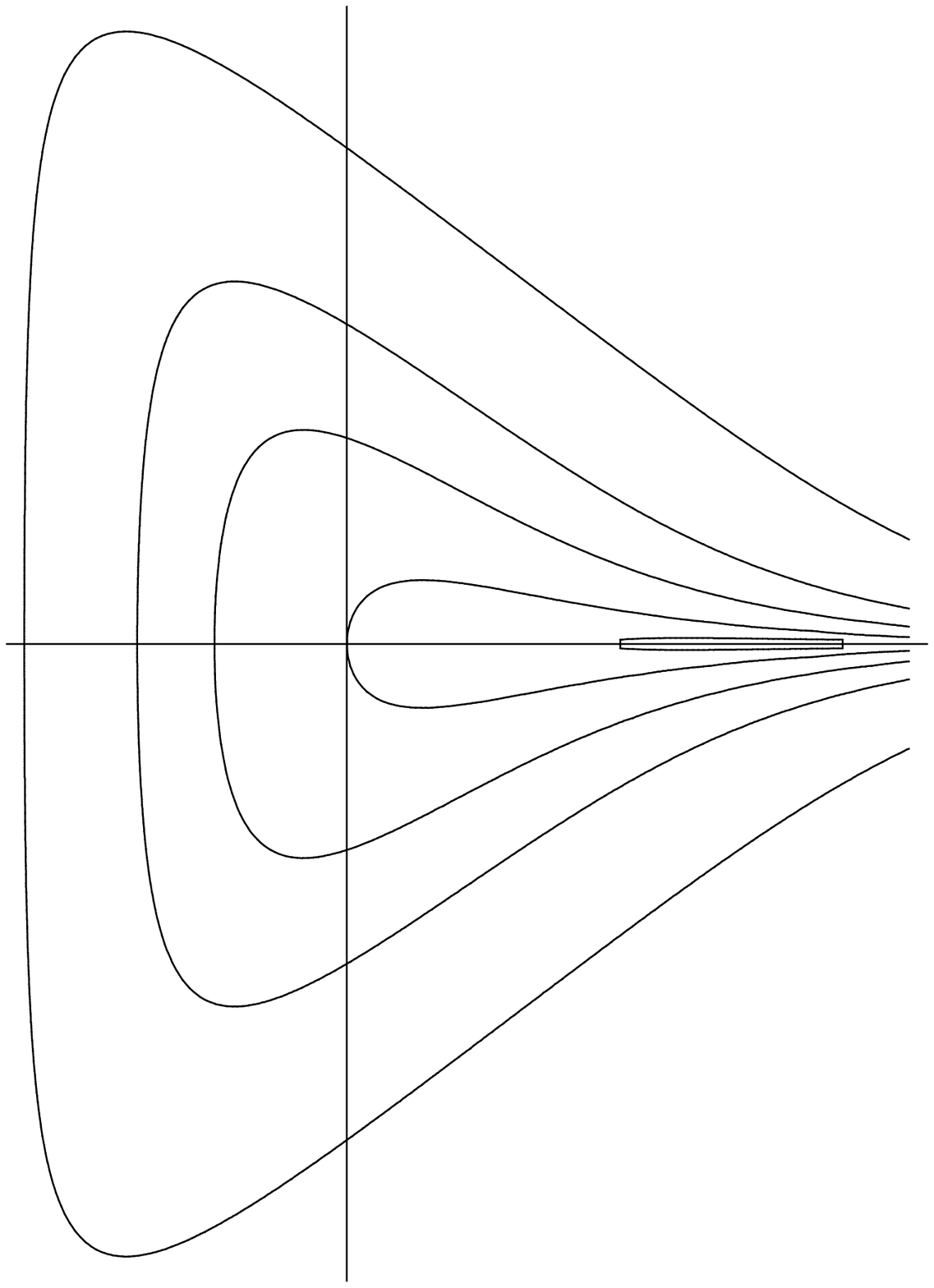}\hspace{1cm}\includegraphics[height=4cm]{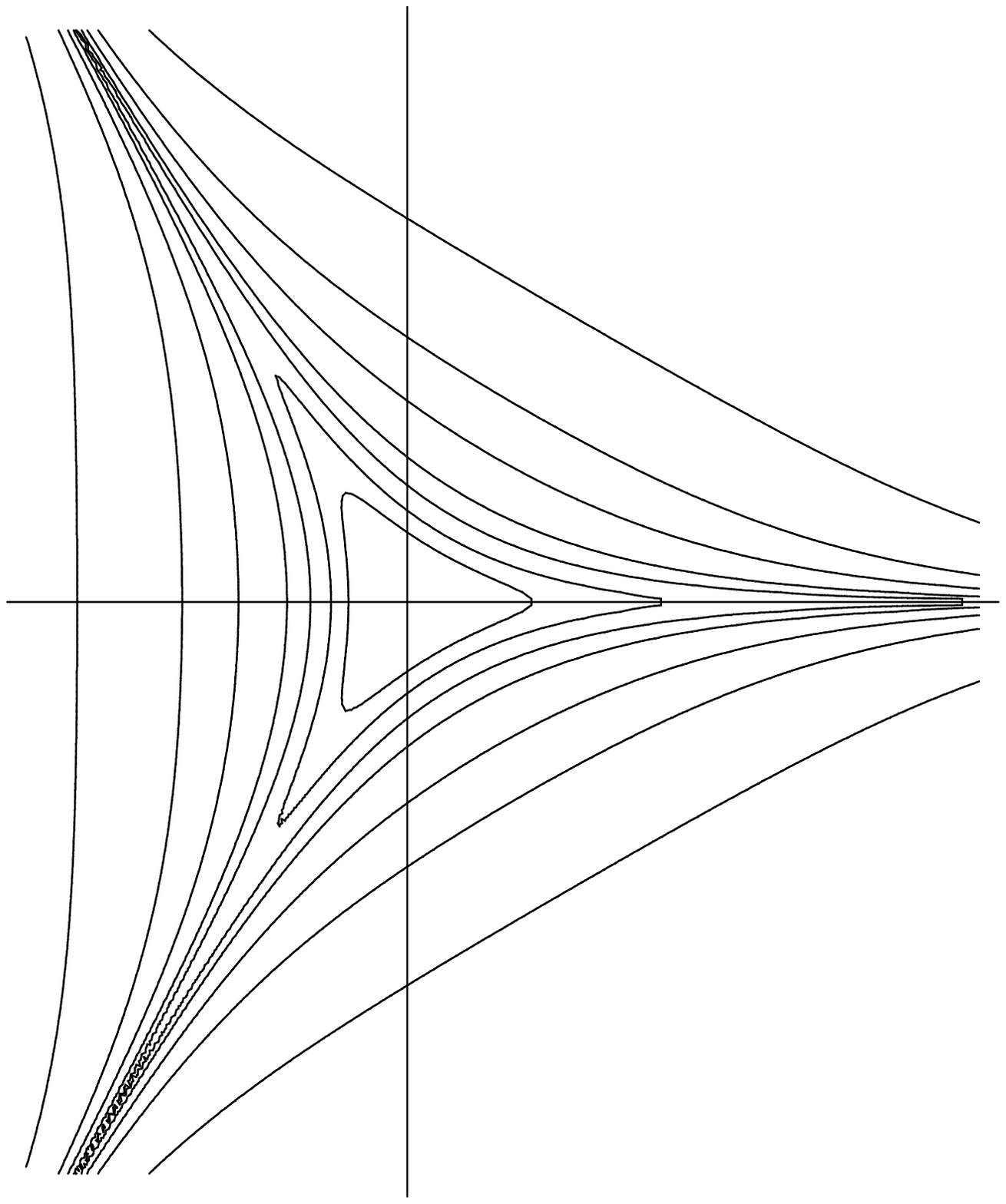}}
\caption{The potentials in the Bianchi type VIII (left) and type XI model (right).}\label{potentials}
\end{figure}

The chaos can be seen more intuitively from a Hamiltonian point of
view. In this picture the Bianchi type VIII and IX models have a
potential of a triangular shape, depicted in Fig. \ref{potentials}.
The ``universe point'' will bounce off the exponentially steep walls and cause the universe to move from one Kasner epoch
into another. This oscillation is  chaotic and is an
attractor set for most matter configurations. 

\section{Chaos in the classical magnetic Bianchi type I}
The presence of cosmic magnetic fields tend to mimic the behaviour in
the vacuum type VIII and IX models. In the type VIII and IX cases the
three-curvature gives rise to the walls seen in
Fig. \ref{potentials}. The magnetic fields give rise to similar walls
and a similar chaotic behaviour results. 

In particular, using the dynamical systems approach, LeBlanc
\cite{LeBlanc} investigated the Bianchi type I with a magnetic field and a perfect
fluid.  The system has no equilibrium points which act
as past attractors, but there
exists a compact subset of phase space which generates chaotic oscillations on
the Kasner circle. This set was conjectured to be an attractor
into the past for generic models. 
There exist more complicated models having chaotic behaviour, but due
to the simpleness of the magnetic Bianchi type I model it can serve as
an effective and simple model to study this type of behaviour. 

\section{Brane-worlds}

Based on investigations of brane-worlds with isotropic fluids it has been suggested by some authors
\cite{MSS:2001,Campos:2001pa,Toporensky:2001hi,Coley1,Coley2} that brane-worlds have an
isotropic singularity\footnote{This would have some interesting
consequences for the Weyl curvature
conjecture \cite{penrose,wcc1,wcc2,wcc3}.} and thus have no chaotic
behaviour. However, this is
an oversimplification of the early time behaviour of
brane-worlds.
Anisotropic stresses are inevitable in anisotropic spacetimes at early
times because of the presence of collisionless gravitons,
collisionless asymptotically-free particles, and electric and magnetic
fields. These anisotropic stresses can arise because of intrinsic anisotropic
stresses on the brane or via the induced graviton stresses from the
bulk. Here we will consider the simplest example: a pure magnetic
field on a flat anisotropic brane with Bianchi type I geometry. 

The evolution equations on the brane are as
follows\cite{Maartens:2001jx,Maartens:2000fg}.
The Friedmann equation,
\begin{eqnarray}
H^2&=&\frac \Lambda 3+\frac 16\sigma ^{\mu \nu }\sigma _{\mu \nu }-\frac 16%
{}^{(3)}\mathcal{R}+\frac{\kappa ^2}3 \rho \nonumber \\ &+&\frac{\kappa ^2}{6\lambda}\left[ \rho^2  -\frac 3{2}\pi _{\mu \nu }\pi ^{\mu \nu }\right]
+\frac{2%
\mathcal{U}}{\kappa ^2\lambda },  \label{constraint}
\end{eqnarray}
the shear propagation equations,
\begin{eqnarray}
\dot{\sigma}_{\langle\mu \nu \rangle}+\Theta \sigma _{\mu \nu }&=&\kappa ^2\pi
_{\mu \nu }-{}^{(3)}\mathcal{R}_{\langle\mu \nu \rangle}\nonumber \\  &+& %
\frac{\kappa ^2}{2\lambda }\left[ -(\rho +3p)\pi _{\mu \nu }+\pi _{\alpha
\langle\mu }\pi _{~\nu \rangle}^\alpha \right] +\frac 6{\kappa ^2\lambda }\mathcal{P}%
_{\mu \nu }, \label{CMeq}
\end{eqnarray}
Raychaudhuri's equation ($\Theta=3H$),
\begin{eqnarray}
&&\dot{\Theta}+\frac 13\Theta ^2+\sigma ^{\mu \nu }\sigma _{\mu \nu }+\frac
12\kappa ^2(\rho +3p)-\Lambda =  \nonumber \\
&&-\frac 1{2\lambda \kappa ^2}\left[ \kappa ^4(2\rho ^2+3\rho p)+12\mathcal{U%
}\right],
\end{eqnarray}
the dark energy propagation equation,
\begin{eqnarray}
&&\dot{\mathcal{U}}+\frac 43\Theta \mathcal{U}+\sigma ^{\mu \nu }\mathcal{P}%
_{\mu \nu }=  \nonumber \\
&&\frac{\kappa ^4}{12}\left[ 3\pi ^{\mu \nu }\dot{\pi}_{\mu \nu }+3(\rho
+p)\sigma ^{\mu \nu }\pi _{\mu \nu }+\Theta \pi ^{\mu \nu }\pi _{\mu \nu
}-\sigma ^{\mu \nu }\pi _{\alpha \mu }\pi _\nu ^{~\alpha }\right].
\end{eqnarray}
Here, $H$ is the Hubble parameter; $\sigma_{\mu\nu}$ is the shear
tensor; ${}^{(3)}\mathcal{R}_{\mu\nu}$ is the 3-curvature; $\rho$ the
energy-density; $p$ the isotropic pressure; $\pi_{\mu\nu}$ the
anisotropic stress tensor; $\mathcal{U}$ is the nonlocal dark energy;
and $\mathcal{P}_{\mu\nu}$ is the nonlocal bulk graviton
stress tensor.

Note that there are no propagation equations for the nonlocal bulk graviton
stress tensor, $\mathcal{P}_{\mu\nu}$ (see also \cite{Deruelle}).

The effect from the anisotropic stresses on the isotropic initial
singularity can be seen by the following considerations (following \cite{Barrow:1997sy,bm2}). We assume
that the we are close to a FRW universe, and hence $H, \rho\propto
t^{-1}$ (note that in the flat isotropic limit we have $H^2\propto
\rho^2$ for brane-world models). For simplicity, we further  assume that the stress tensor is on the
general form\footnote{This includes
the magnetic field case.} $\pi_{\mu\nu}=C_{\mu\nu}\rho_{r}$  where $C_{\mu\nu}$ is a
constant trace-free matrix, and $\rho_r$ is the energy-density of a
radiation type of fluid causing the anisotropic stresses (hence, we 
assume that the isotropic pressure $p_r$ is $p_r=\rho_r/3$). Including
also an isotropic fluid with equation of state $p_i=(\gamma-1)\rho_i$,
we note that the isotropic past singularity is unstable due to the
anisotropic stresses whenever $\gamma \leq 4/3$. Thus this simple
investigation leads to the conclusion: \emph{For brane-worlds with
isotropic perfect fluids with
$\gamma\leq 4/3$ the initial singularity cannot be isotropic in the
presence of stresses of type  $\pi_{\mu\nu}=C_{\mu\nu}\rho_{r}$, and
thus for magnetic fields in particular.} 
The anisotropic stresses will therefore be very important for the
early time behaviour of brane-world models. The initial singularity is
\emph{matter dominated} in contrast to the shear dominated singularity
in GR. 

This makes us wonder whether the inclusion of a magnetic field (or more
general types of stresses) will make the chaos come back to the
Bianchi type I model. More generally we can ask ourself: What is
the nature of the singularity for a magnetic brane-world?

To investigate this  we used the dynamical systems
approach and considered a Bianchi type I brane-world \cite{BH} with
 an isotropic perfect fluid, $p=(\gamma-1)\rho$;  a magnetic field, $\pi_{\mu\nu}=-B_{\mu}B_{\nu}+\frac
13B^2h_{\mu\nu}$ ; and $\mathcal{P}_{\mu\nu}=0$ in order for
the equations to close. 

By finding all the equilibrium points and by investigating their
stability, we found the following two (one for $\gamma\leq 4/3$) \textbf{past
attractors}:
\begin{enumerate}
\item \underline{Isotropic FRW}: \\ 
Scale factor: 
 $a \propto t^{\frac {1}{3\gamma}} $ \\
Past attractor for $\gamma > \frac 43$. 

\item \underline{An anisotropic solution}: \\
A past attractor with metric
\beq{
ds^2=-dt^2+t^{-\frac 1{68}(\sqrt{3345}-43)}dx^2+t^{\frac 12}(dy^2+dz^2).
}\end{eqnarray}
The magnetic field and the dark-energy term diverge like
 $
B^2\propto t^{-1}$ and $\mathcal{U}\propto t^{-2}$, respectively. This is an
\textbf{attractor for all values of $\gamma$}. For more details about
the equilibrium points and their nature, consult \cite{BH}. 
\end{enumerate}

The presence of an attractor solution in the magnetic Bianchi type I
model may indicate that there are \textbf{no chaos} in the brane-world
models. 

We now see how the chaos is avoided in the brane-world model. The
brane-world is very sensitive to the matter it contains which
actually dictates the initial singularity almost
entirely. The initial 
singularity does by no means need to be isotropic for these models, as
a matter of fact, it can equally well be a matter/dark-energy dominated anisotropic singularity. The presence of the dark-energy
term indicates that the Weyl tensor in the bulk is non-zero and could
therefore be 
important initially. This may be a signal that the assumption that the
nonlocal bulk graviton stress tensor is zero is an artificial and
unnatural one. However, work done by others \cite{SST} indicates that
in the absence of intrinsic stresses the past asymptotes does not
change significantly with the inclusion of a non-zero
$\mathcal{P}_{\mu\nu}$. This goes in favour of our assumption, but
further work is needed on this issue.

\section{Conclusion}
We have seen how anisotropic stresses can alter the behaviour at early
times for brane-world models. At the high energies which occur at 
early times in the evolution of the universe, the brane-world  is very
 dependent on the  physical matter content. This is in contrast to
the classical general relativistic case, where the early times is
mostly dominated by the shear. In particular, we saw how the inclusion
of anisotropic stresses could
change the behaviour of the magnetic Bianchi type I model quite drastically; the
classical model has a chaotic past, while the brane-world has not. 

It should be noted that even though we have ruled out a chaotic behaviour in the past of the
mixmaster type, we have not ruled out other types of chaos for
brane-world models. For example, the chaos in models with Yang-Mills fields at
late times \cite{jbjl}, is not ruled out by our analysis. Further
work is needed on this issue. 

Notwithstanding that the brane-world model is very unlikely to have
chaos is the past, the
initial singularity could still be very anisotropic. This has later
been suggested by others \cite{BD}. More specifically, the initial
singularity in the magnetic
Bianchi type I brane-world  can only be
isotropic if the universe contains an isotropic fluid  stiffer
than radiation (thus if $\gamma>4/3$). 

\begin{acknowledgements}
The author would like to thank the local organising committee for
making this talk possible and to A.A. Coley to useful comments and suggestions.. 
Funding from the Research Council of Norway and a travel grant from
Churchill College, are also gratefully acknowledged.
\end{acknowledgements}

\end{article}
\end{document}